# Efficient single-photon pair generation by spontaneous parametric down-conversion in nonlinear plasmonic metasurfaces

Boyuan Jin, Dhananjay Mishra, and Christos Argyropoulos*
Department of Electrical and Computer Engineering, University of Nebraska-Lincoln,
Lincoln, Nebraska 68588, USA
*christos.argyropoulos@unl.edu

**Abstract**

*Spontaneous parametric down-conversion (SPDC) is one of the most versatile nonlinear optical techniques for the generation of entangled and correlated single-photon pairs. However, it suffers from very poor efficiency leading to extremely weak photon generation rates. Here we propose a plasmonic metasurface design based on silver nanostripes combined with a bulk lithium niobate ($LiNbO_3$) crystal to realize a new scalable, ultrathin, and efficient SPDC source. By coinciding fundamental and higher order resonances of the metasurface with the generated signal and idler frequencies, respectively, the electric field in the nonlinear media is significantly boosted. This leads to a substantially enhancement in the SPDC process which, subsequently, by using the quantum-classical correspondence principle, translates to very high photon-pair generation rates. The emitted radiation is highly directional and perpendicular to the metasurface on the contrary to relevant dielectric structures. The incorporation of circular polarized excitation further increases the photon-pair generation efficiency. The presented work will lead to the design of new efficient ultrathin SPDC single-photon nanophotonic sources working at room temperature that are expected to be critical components in free-space quantum optical communications. In a more general context, our findings can find various applications in the emerging field of quantum plasmonics.*

# 1. Introduction

The field of quantum photonics is poised to have a broad range of emerging applications spanning ultrafast speed and robust security quantum communications,[1-8] rapid and accurate quantum computations,[9-16] and high resolution metrology, imaging, and sensing.[17-20] The generation of high-quality entangled single-photon pairs by nonclassical light sources is an essential building block for the majority of these applications and has been a long-sought goal in modern quantum optical technologies.[21, 22] The process of spontaneous parametric down-conversion (SPDC), alternative known as parametric fluorescence, in a nonlinear crystal is one of the most versatile techniques for the generation of entangled and correlated single-photon pairs.[23-30] However, as a typical spontaneous quantum nonlinear process, it suffers from extremely poor efficiency due to the very weak nature of nonlinear quantum optical processes.[31] Recently, extensive efforts have been devoted to boost this effect by utilizing conventional designs based on nonlinear crystals,[32] waveguides,[33, 34] photonic crystals,[35] and ring resonators[36, 37] that are usually bulky or unsuitable for free space applications.

On the contrary, metasurfaces[8] that can sustain subwavelength confined gap-plasmons are promising novel ultrathin designs to realize saleable and efficient SPDC sources, as they can tightly confine the incident optical power locally in ultrasmall nanoscale volumes resulting to enhanced nonlinear optical effects.[38-44] Thus, the SPDC photon-pair generation rate process, which is proportional to the local pump intensity and induced field enhancement, is expected to be substantially improved by these compact plasmonic configurations. Recently, the SPDC process has been enhanced by using a single dielectric cylindrical nanoantenna.[45] The extension of this configuration to antenna arrays is expected to further boost its output signal but remains elusive. In addition, despite the low optical loss of this dielectric nanoantenna, its photon-pair generation rate is limited mainly due to the low field enhancement along its geometry leading to relative weak nonlinear light-matter interactions. Interestingly, most of the configurations presented in the literature, employed to enhance SPDC, use III−V semiconductor materials due to their large second-order nonlinearity.[31] However, it is very challenging for the majority of these designs to get the emitted photon pairs in the normal direction to the sample's surface combined with enhanced single-photon pair generation rate while operating as a free-standing configuration (not waveguide).

In this work, we resolve these problems by designing a new plasmonic nonlinear metasurface to efficiently generate entangled single-photon pairs. The metasurface can sustain a loosely confined gap-plasmon mode and is composed of periodic silver nanostripes placed on top of a homogeneous lithium niobate (LiNbO$_3$) spacer layer with the whole structure terminated by a silver substrate. LiNbO$_3$ has relatively strong second-order nonlinearity which can be of great potential in SPDC applications. However, it is usually used in elongated bulky waveguide configurations exhibiting relatively low SPDC efficiency.[46-49] In the currently proposed structure, the induced localized surface and gap plasmon modes generate extremely strong electric field enhancement around the edges of the silver nanostripes and, in particular, the nanogap region. The bottom silver substrate works as a perfect reflector which creates loosely confined gap plasmons due to the induced Fabry-Pérot resonances in the nanogap that further enhance the electric field in the LiNbO$_3$ nonlinear spacer layer. The generated electric field enhancement hotspots coincide with the metasurface second-order nonlinear materials, mainly LiNbO$_3$ and to a lesser extend the silver-dielectric interface.

To achieve the maximum entangled photon-pair generation rate due to the boosted SPDC process, the metasurface resonances coincide to both generated signal and idler frequencies. Note that the SPDC process is usually complicated and requires random variables to be characterized due to its pure quantum nature. To decrease the computational complexity of the current problem, we simulate the reverse classical nonlinear process, named sum-frequency generation (SFG), and then employ the quantum-classical correspondence relationship to calculate the photon-pair generation rate, in a similar way to various previous works relevant to different SPDC systems.[45, 50, 51] It is demonstrated that the photon-pair generation rate in the proposed nonlinear metasurface is substantially increased compared to the current state-of-the-art designs with equivalent ultrathin profiles operating as free-standing structures (not as waveguides). We thoroughly investigate the factors that affect the obtained efficient SPDC process, such as operation frequency, incident wave polarization, and photon emission angles. The various media composing the proposed metasurface and consisting the main sources of the SPDC process strength are also explored. The proposed new ultrathin SPDC metasurface design works in reflection mode and generates a spatially narrow photon emission beam perpendicular to its surface. It also has high efficiency and can be used in room temperature, which are major advantages compared to alternative solid-state single photon emitters that usually operate in cryogenic temperatures.[22] Moreover, it can easily be made frequency tunable by varying its geometry, making it possible to work for various input and output generated wavelengths. Its versatile and efficient SPDC response is ideal to be used in vital components of various emerging free-space quantum optical devices. The presented generation of high-quality entangled single-photon pairs can find a plethora of interesting applications in the emerging field of quantum plasmonics.

## 2. Nonlinear Metasurface to Boost the SPDC Process

SPDC is generated from the spontaneous parametric amplification of vacuum thermal noise photons in a nonlinear medium when pumped at low enough power to avoid any stimulated photon generation.[30] It can lead to the spontaneous generation of two correlated and entangled photons at the angular frequencies $\omega_s$ and $\omega_i$, respectively, following the annihilation of a pump photon at frequency $\omega_p$. The subscripts $s$, $i$, and $p$ represent the signal, idler, and pump photons, respectively, and the conservation of energy $\omega_p = \omega_s + \omega_i$ is always satisfied. The process is schematically shown in the inset of Fig. 1a. SPDC is restricted from the phase-matching condition (i.e., combined conservation of energy and momentum) which is the most typical limitation for any nonlinear optical process.[31]

The proposed nonlinear plasmonic metasurface design is schematically depicted in Fig. 1. Periodic silver nanostripes are placed on top of a LiNbO$_3$ spacer layer and terminated by a silver substrate. The thickness, width, and periodicity of the silver nanostripes are: $b = 100$ nm, $l = 450$ nm, and $a = 560$ nm, respectively. The edges of the silver nanostripes are rounded with a radius $r = 2$ nm to imitate experimental fabrication imperfections.[39] The thickness of the LiNbO$_3$ layer is chosen to be $h = 306$ nm after optimization with the goal to tune the fundamental and higher order resonances to the involved waves, as it will be shown later. Note that similar thickness LiNbO$_3$ layers have been experimentally verified in various recent papers[52-54] and were recently grown over a metal bottom reflector,[55] similar to the currently proposed configuration. The illumination used to excite the current structure is always a plane wave with electric field parallel to the x-direction or, equivalently, transverse magnetic (TM) polarization. The nanostripe metasurface will be polarization dependent and will not work for the transverse electric (TE) polarization. However, it can become polarization independent without altering its performance if the nanostripes are replaced by an array of three-dimensional (3D) nanocube resonators[43] that are polarization insensitive.

Localized surface plasmons are formed at the metasurface fundamental and higher-order resonances leading to the generation of extremely strong electric fields around the edges of the silver nanostripes and in the nanogap. Furthermore, the bottom silver substrate operates as a reflector, forming a loosely confined gap-plasmon mode due to an additional Fabry-Pérot resonance that further increases the electric field in the LiNbO$_3$ spacer layer. The proposed metasurface works only in reflection mode and the transmission is zero due to the thick bottom silver substrate. The reflectance spectrum in the linear operation of the proposed metasurface is computed under normal incident illumination and shown in Fig. 2a. The obtained reflectance dips suggest that the metasurface has two resonance modes in the near-infrared (IR) spectrum, both accompanied by large field enhancement $|E/E_0|$, where E is the local electric field, $E_0$ is the amplitude of the incident field, and $|E/E_0|_{max}$ is the maximum field enhancement. The electric field is substantially boosted in the LiNbO$_3$ layer and along the silver nanostripe and

substrate surfaces, as depicted in Fig. 2b. These consist the main second-order nonlinear materials of the proposed metasurface. It is well known that the generally weak nonlinear effects need to be excited by intense optical input intensities.[56, 57] Hence, the local electric field enhancement achieved by the current plasmonic metasurface is critical for improving its efficiency.[41, 58] This principle is general and can be applied to both stimulated and spontaneous nonlinear processes, including the currently studied SPDC. As a result, it is expected that if we coincide $\lambda_i$ and $\lambda_s$ with the fundamental (1479 nm) and higher-order (1254 nm) resonances, respectively, of the current nonlinear metasurface design the efficiency of the SPDC process will be substantially improved.

Structures designed to enhance classical second-order nonlinear processes can also be used to enhance the purely quantum optical SPDC process. Indeed, it has been proven that there is a quantum-classical correspondence between the SPDC and SFG processes, respectively, based on the Green's function formalism.[45, 50, 51] More specifically, the number of photon pairs generated through SPDC are proportional to the SFG conversion efficiency, while the pump waves ($\omega_1$ and $\omega_2$) and the generated sum-frequency wave ($\omega_{SF} = \omega_1 + \omega_2$) of the classical SFG process propagate in the opposite directions to the signal ($\omega_s$), idler ($\omega_i$), and pump ($\omega_p$) photons in the SPDC process.[45, 50, 51] The SFG process is schematically shown in the inset of Fig. 1b. The quantum-classical correspondence concept is also demonstrated in Fig. 1 and can be applied to any system that is reciprocal in the linear regime, allowing to predict the efficiency of the SPDC generated entangled photon pairs by studying the classical SFG nonlinear process under reversed direction interacting waves. In this case, the entangled photon-pair generation rate due to the quantum SPDC process is given by the formula:[45]

$$\frac{dN_{pair}}{dt} = 2\pi c \frac{\lambda_p^4}{\lambda_s^5 \lambda_i^3} \frac{I_p \Delta \lambda_s}{A} \eta_{SFG},$$ (1)

where $\lambda_p$, $\lambda_i$, and $\lambda_s$ are the wavelengths of the pump, idler, and signal photons, respectively, $c$ is the light speed in vacuum, $I_p$ is the SPDC incident pump wave intensity, and $A$ is the illumination area along the metasurface. The photon-pair rate is proportional to the bandwidth of the generated signal $\Delta \lambda_s$. Since the SPDC is a spontaneous effect, all the combinations of $\lambda_i$ and $\lambda_s$ that satisfy the relation $\omega_p = \omega_s + \omega_i$ will be produced from noise photons generated by quantum vacuum fluctuations.[59] Therefore, $\Delta \lambda_s$ in Eq. (1) is determined by the spectral response of the structure and, even more importantly, the bandwidth of the detector. Finally, $\eta_{SFG} = P_{SF} / I_1 / I_2$ represents the efficiency of the corresponding classical SFG nonlinear process, where $P_{SF}$ is the power outflow of the generated sum-frequency wave due to the SFG process, and $I_1$ and $I_2$ are the incident intensities of the SFG pump waves. Note that strong field enhancement is not required at the pump frequency of the SPDC process to increase the single-photon pair generation rate efficiency, as it is predicted by Eq. (1). It can also be detrimental to the SPDC process because it can lead to stimulated photon generation.[30] The measured $P_{SF}$ always depends on the illumination area $A$ along the sample. However, we normalize the calculated photon-pair generation rate to the sample's surface $A$ by dividing this area on the right side of Eq. (1). In addition, the photon-pair generation rate is proportional to the incident pump intensity $I_p$ and the efficiency of the corresponding SFG process $\eta_{SFG}$. However, as we will show later, $\eta_{SFG}$ is independent of the incident intensities $I_1$ and $I_2$. To improve the $\eta_{SFG}$ and thus enhance the SPDC process, the nanophotonic structure of the sample needs to be carefully designed and optimized.

We avoid the complicated quantum calculations involved in the SPDC process by utilizing the quantum-classical correspondence and studying the easier to model classical SFG process. Hence, our study is simplified to the investigation of a nonlinear plasmonic metasurface design that can exhibit exceptionally strong SFG response. The SFG conversion efficiency is calculated by using nonlinear full-wave simulations based on the finite element method (FEM) software COMSOL Multiphysics. These electromagnetic simulations are not trivial, since the linear Maxwell's equations need to be substantially modified in COMSOL by introducing a nonlinear polarizability to model the SFG nonlinear process in the nanoscale. More details about the linear and nonlinear modeling are provided in the Supplementary Material.[60] Periodic boundary conditions are used and only one unit-cell of the nanostripe plasmonic metasurface is modeled. Moreover, considering that the length of the silver nanostripes is much longer than the currently used near-IR wavelengths, the structure is modeled as two-dimensional (2D), which substantially decreases the computational speed and memory burden of the computationally demanding nonlinear simulations. It should also be noted that the linear permittivities of silver and LiNbO$_3$ used in the simulations have practical values taken from experiment data.[61, 62]

LiNbO$_3$ is an emerging and currently widely used bulk nonlinear material due to its relatively strong second-order nonlinearity.[46-49] The three orthogonal components of the induced anisotropic nonlinear polarizability in the SFG process are given by:[45, 63]

$$P_x^{NL} = 2\varepsilon_0 \left[ d_{33} E_{1x} E_{2x} + d_{31}(E_{1y} E_{2y} + E_{1z} E_{2z}) \right], \tag{2}$$

$$P_y^{NL} = 2\varepsilon_0 \left[ d_{31}(E_{1y}E_{2x} + E_{1x}E_{2y}) + d_{22}(E_{1y}E_{2y} - E_{1z}E_{2z}) \right], \quad (3)$$

$$P_z^{NL} = 2\varepsilon_0 \left[ d_{31}(E_{1z}E_{2x} + E_{1x}E_{2z}) - d_{22}(E_{1y}E_{2z} + E_{1z}E_{2y}) \right], \quad (4)$$

where $\varepsilon_0$ is the permittivity of free space, and $d_{31} = 5.95$ pm/V, $d_{33} = 34.4$ pm/V, and $d_{22} = 3.07$ pm/V are the non-zero elements of the LiNbO$_3$ crystal anisotropic second-order nonlinear susceptibilities.[31, 64] The subscripts *x, y, z* in Eqs. (2)-(4) represent the corresponding components of the induced nonlinear polarizability and electric field along different axes. Due to the non-centrosymmetric crystalline structure of LiNbO$_3$, its second-order nonlinear response is anisotropic and relatively high. This is on the contrary to centrosymmetric crystals that, in principle, cannot have a second-order nonlinear response.[31] In order to utilize the predominant $d_{33}$ nonlinear LiNbO$_3$ susceptibility in the current scenario of x-polarized incident waves, the crystalline optical axis of LiNbO$_3$ in our design is aligned along the x-direction. In III−V semiconductor materials, such as Gallium Arsenide (GaAs) and Aluminum Gallium Arsenide (AlGaAs), which are also non-centrosymmetric but have stronger second-order susceptibilities, the SFG frequency mixing process is mainly pronounced when the polarization directions of the two input waves are orthogonal.[45, 65] This specific property of III−V semiconductor nanostructures makes them challenging to generate sum-frequency radiation collinear to the incident waves.[45, 63, 66] Since normal incident inputs are preferable in most applications, the generated sum-frequency radiation and, consequently, SPDC entangled single-photon pairs from III−V semiconductor materials usually cannot be perpendicular to the sample surface. On the contrary, LiNbO$_3$ achieves the highest SFG conversion efficiency when the inputs are parallel polarized. Thus, the generated sum-frequency wave can be easily emitted perpendicular to the currently proposed nonlinear plasmonic metasurface. This is another major advantage of the currently proposed free-standing efficient SPDC source compared to various relevant designs based on elongated photonic waveguide structures. In addition, the presented plasmonic metasurface can also compensate the relatively small (compared to semiconductors) nonlinear susceptibilities of LiNbO$_3$. As we will show later, the photon-pair generation rate in our proposed plasmonic metasurface is several times larger compared to AlGaAs nanoantennas[45] despite the larger nonlinear susceptibility of AlGaAs compared to the currently used LiNbO$_3$.

Note that enhanced second-order nonlinear effects can also be generated at the metal-dielectric interface,[67-70] in addition to the bulk LiNbO$_3$ nonlinear material. This surface second-order nonlinearity originates from the asymmetry of the neighboring different atoms along the interface. Due to the ultrathin interface region, this surface nonlinearity is more accurate and convenient to be modeled by a boundary condition, similar to 2D nonlinear materials, such as graphene.[71-73] The tangential component of the surface SFG polarizability $P_{s\|}^{NL}$ gives rise to a surface current density $J_{s\|}^{NL}$, but is one-order of magnitude smaller than the normal component $P_{s\perp}^{NL}$.[74, 75] Therefore, $P_{s\|}^{NL}$ is usually assumed to be negligible for simplicity. Here, the subscript $s$ represents "surface". On the other hand, $\mathbf{P}_{s\perp}^{NL} = 2\varepsilon_0 \chi_{s\perp}^{(2)} E_{1\perp} E_{2\perp} \hat{\mathbf{r}}_\perp$ results in a discontinuity of $E_\|$ along the interface,[76] where $\hat{\mathbf{r}}_\perp$ is the unit vector normal to the interface, and $E_{1\perp}$ and $E_{2\perp}$ are the normal components in the metal region of the two input fields, respectively. This surface source generated by $P_{s\perp}^{NL}$ can be modeled as an equivalent surface magnetic current density $\mathbf{J}_{m,s}^{NL} = \hat{\mathbf{r}}_\perp \times (\nabla_\| P_{s\perp}^{NL})/\varepsilon'$,[76] where $\varepsilon'$ is the permittivity of the adjacent dielectric material, and $\nabla_\|$ is the gradient operator on the tangential direction. Note that $\chi_{s\perp}^{(2)}$ is the overall surface component of the bulk nonlinear susceptibility $\chi_{b}^{(2)}$ due to the minor penetration depth of light in the metal. Theoretically, $\chi_{s\perp}^{(2)}$ can be expressed as $\chi_{s\perp}^{(2)} = h_{eff} \chi_{b}^{(2)}$, where $h_{eff}$ is the effective thickness of the metal-dielectric interface.[77] However, the value of $\chi_{s\perp}^{(2)}$ can also be derived from experimental measurements without knowing the values of $\chi_{b}^{(2)}$ and $h_{eff}$.[78, 79] In this work, we use $\chi_{s\perp,Ag}^{(2)} = 1.59 \times 10^{-18}$ m$^2$/V for silver according to relevant reported experimental data.[74, 77, 80] This value is extremely low, especially compared to lithium niobate, since silver is centrosymmetric material, and does not play a substantial role in the SFG process, as it will be shown in the next section.

## 3. Significantly Boosted SFG Process

We characterize the SFG process strength by computing the conversion efficiency $CE_{SFG} = P_{SF} / (P_1 + P_2)$, where $P_{SF}$ is the power outflow of the generated sum-frequency wave, and $(P_1 + P_2)$ is the total input power from both pump waves.[81, 82] As seen by the red dashed line in Fig. 2a, the electric field is greatly boosted at the resonant wavelengths of the proposed plasmonic metasurface. When the input pump waves operate at $\lambda_1$ and $\lambda_2$ wavelengths that are chosen to vary close to the metasurface two resonance modes (1254 nm and 1479 nm, respectively), the computed SFG conversion efficiency is shown in Fig. 3a and takes relatively high values. The intensities of the incident waves are both fixed to the moderate value of 100 MW/cm$^2$, which ensures that the sample will not be affected by detrimental thermal effects and eventually destroyed (melt) by heating. Pulsed lasers can be used to improve the heat dissipation. Interestingly, even higher input intensity values were used before in other nonlinear experiments based on similar plasmonic metasurfaces without causing damage to the sample.[39]

Moreover, the undepleted-pump approximation is always adopted in all SFG simulations, since the generated wave is much weaker compared to the input waves.

The maximum conversion efficiency can exceed $1.8 \times 10^{-6}$ in the case of $\lambda_1 = 1257$ nm and $\lambda_2 = 1477$ nm, leading to a generated sum-frequency wave at the visible wavelength $\lambda_{SF} = 679$ nm. The SFG conversion efficiency also depends on the input intensity, as shown in Fig. 3b, when plotted at the same wavelengths, suggesting that even higher conversion efficiency values can be achieved just by increasing the input power. The proposed metasurface realizes high conversion efficiency that is comparable (or even improved) to different previously reported structures with dramatically enhanced second-order nonlinearity.[83, 84] We also investigate the distinct contribution of each nonlinear material involved in the metasurface design. As depicted by the red dashed line in Fig. 3b, the reduction in the conversion efficiency is minor to non-existent if the nonlinearity of the silver-dielectric interface is neglected ($\chi_{sAg}^{(2)} = 0$). Although the field enhancement on the metal surface is extremely large, it is derived that the dominant contribution in the second-order nonlinear process is mainly from the LiNbO$_3$ layer. When the silver nanostripes are removed from the metasurface, the conversion efficiency is substantially decreased by five orders of magnitude, as illustrated by the blue dashed line in Fig. 3b. Hence, the nanocavities formed by the geometry of the presented structure are crucial to achieve substantially enhanced nonlinear operation.

As discussed before, nonlinear metasurfaces made of III−V semiconductor materials are difficult to radiate the sum-frequency generated wave perpendicular to the sample surface. To investigate the directivity of our proposed metasurface, Fig. 4 shows the computed far field radiation pattern of one unit cell, where $\theta_1$ and $\theta_2$ are the incident angles of the two input pump waves, respectively. Under normal ($\theta_1 = \theta_2 = 0°$) or symmetrically oblique ($\theta_1 = -\theta_2$) illumination, most of the generated power of the sum-frequency wave is directionally reflected to free space normal to the metasurface. It is also clear that the nonlinearly generated radiation decreases under small oblique incident angle illumination. More details about the far field calculations are provided in the Supplementary Material.[60]

Due to the anisotropic nature of the $\chi^{(2)}$ tensor in the involved nonlinear materials and the structure's polarization sensitivity, the generated sum-frequency radiation will be affected by the input wave polarization. In the case of linear polarization, we define the orthogonal- (O) and parallel- (P) polarizations when the incident plane wave electric field $\mathbf{E}_0$ is orthogonal (TM polarization) or parallel (TE polarization) to the silver nanostripes. The SFG conversion efficiency is computed and presented in Fig. 5a under different combinations of the incident polarizations including O or P, and, additionally, right- and left-handed circular polarizations (RCP and LCP, respectively). The input waves are always normally incident to the metasurface with intensities $I_1 = I_2 = 100$ MW/cm$^2$. When either one of the two incident pump waves is P-polarized, it can be seen in Fig. 5a that the generated sum-frequency power is extremely low and close to zero. This is because the metasurface resonance performance is polarization sensitive, i.e., it cannot resonate under P-polarization, and thus the field cannot be enhanced in the metasurface nanogap. In any other combination, the radiation at the sum-frequency is strong. The conversion efficiency shown in Fig. 5a is computed only for the O-polarized sum-frequency power outflow ($0.5 \times \text{Re}[-H_{SF,y}^* E_{SF,x}]$), since the y-component of the sum-frequency electric field $E_{SF,y}$ is much lower and can be neglected for any polarized excitation. This suggests that the generated sum-frequency wave has purely linear polarization with electric field along the x-direction (O-polarized). Interestingly, the angular-dependent polarization state of the sum-frequency generated wave can be derived by the SFG far-field radiation pattern shown in Fig. 4. The power of the sum-frequency nonlinearly generated wave is maximum at normal emission angles and abruptly decreases as the emission angles deviate from normal. The pattern shown in Fig. 4 has always the same shape even in the case of different angle incident wave illuminations. In addition, the y-component of the electric field of the sum-frequency nonlinearly generated wave is always very low compared to the x-component for different angles of emission even off the normal emission angle. As a result, the generated sum-frequency wave will be linear polarized with electric field along the x-direction which however will have much lower amplitude as the emission angles increase and deviate compared to normal. This is mainly due to the resonance mode that is angle-dependent and can be perfectly excited only for normal incident waves.

It is interesting that circular polarization can generate even stronger SFG, as clearly demonstrated in Fig. 5a. In fact, the relative SFG strength between linear and circular polarized input pump waves depends on the operation frequency. For example, Fig. 5b shows the maximum field enhancement in the LiNbO$_3$ layer as a function of $\lambda_1$ for different polarizations operating in the linear regime. It can be seen that the field enhancement for both circular polarizations at the resonance is larger compared to the linear polarization, since magnetic modes are excited in the nanogap region that couple more efficiently to circular polarized incident waves.[85-88] In the Supplementary Material,[60] we also compute the linear reflectance spectra when the plasmonic metasurface is excited by circular polarized waves, where it is shown in Fig. S1 that both reflection dips are deeper compared to those under linear polarized excitation. In addition, we clearly demonstrate the magnetic nature of the excited resonance modes by computing the magnetic field enhancement distributions in Fig. S2.[60] This suggests that the SFG under RCP and/or LCP polarized pump input beams will be stronger at the resonance wavelength compared to the linearly polarized pump waves. On the contrary, linear O-polarized pump waves can generate stronger SFG at off-resonance frequencies, where they can produce larger electric field than the RCP and LCP inputs, as shown in Fig. 5b. The field enhancement for the P-polarized input waves is also shown by the red solid line in Fig. 5b and is always extremely small and less than one when varying $\lambda_1$. As was mentioned before, plasmonic resonances cannot be excited under P-polarization with the current nanostripe-based metasurface, resulting in negligible SFG efficiency under this polarization. However, we should note that the relation $|E/E_0|_{max}$ is only a coarse indicator to characterize the SFG strength. To be more accurate, the SFG conversion efficiency is proportional to $\iiint P^{NL} dV$, where $P^{NL}$ is given by Eqs. (2)-(4) presented in the previous section that include the LiNbO$_3$ anisotropy. To conclude, the SFG nonlinear process is substantially enhanced by the current nonlinear plasmonic metasurface when illuminated by either O- or circular polarized pump waves.

## 4. Highly Efficient Photon-Pair Generation due to Enhanced SPDC

Based on the SFG simulations presented in the previous section, we extract the power outflow of the generated sum frequency wave along the unit cell of our proposed metasurface, named $P_{SF}$, and compute the SFG efficiency $\eta_{SFG}$ by dividing $P_{SF}$ with the incident intensities $I_1$ and $I_2$ of the pump waves. Unlike the SFG conversion efficiency $(CE_{SFG})$ used before, the $\eta_{SFG}$ metric is independent of the incident intensities and can only be enhanced by the nanophotonic structure. The photon-pair generation rate in the reversed SPDC process is computed by using the quantum-classical correspondence defined before by Eq. 1. We assume that the metasurface is illuminated uniformly by a perpendicularly incident plane wave, also known as the pump ($\lambda_p$). In Eq. 1, we assume that the sample area $A$ is approximately mm$^2$ scale. Thus, the derived photon-pair generation rate is for a 1 mm$^2$ metasurface and has units of Hz/mm$^2$. In addition, the metasurface photon-pair generation rate is proportional to the bandwidth of the generated signal and the input pump intensity. Therefore, to make the comparison more convenient, we normalize the computed photon-pair generation rate to the signal bandwidth $\Delta\lambda_s = 1$ nm and the used pump intensity $I_p = 1$ W/cm$^2$. The pump intensity $I_p = 1$ W/cm$^2$ used in our computations is extremely low compared to typical nonlinear applications to avoid any stimulated photon generation. Moreover, the excitation electric field enhancement in the nanogap at the pump signal ($\lambda_p$) has much lower value compared to the electric field enhancement at the metasurface resonances shown in Fig. 2(a). The computed photon-pair generation rate of our proposed metasurface as a function of the signal and idler wavelengths is shown in Fig. 6a. When we vary both signal and idler waves close to the resonant wavelengths of the proposed plasmonic metasurface, the photon-pair generation rate can reach to a very high value of approximately 1400 Hz/mm$^2$. Currently, the best state-of-the-art ultrathin designs based on alternative nanoantenna or metasurface structures can only generate photon pairs with an estimated rate of approximately 400 Hz/mm$^2$ or 72 Hz/mm$^2$, respectively.[45, 89] Hence, the proposed nonlinear plasmonic metasurface can significantly increase the photon-pair generation rate. This rate can be even further increased if the LiNbO$_3$ spacer layer becomes thinner, which will lead to stronger electric field enhancement and, as a result, increased nonlinear light-matter interactions. According to the quantum-classical correspondence, the extremely high photon-pair generation rate occurs for the following visible and near-IR wavelengths: $\lambda_p = 679$ nm, $\lambda_s = 1257$ nm, and $\lambda_i = 1477$ nm. The computed photon-pair generation rate as a function of the signal wavelength is also shown in Fig. 6b, where the pump is fixed to $\lambda_p = 679$ nm. This plot proves that the enhanced photon generation is indeed due to the resonances excited by the metasurface. The maximum photon-pair generation rate is obtained at $\lambda_s = 1257$ nm, which coincides with the metasurface's higher-order resonance.

The quantum-classical correspondence also implies that the input pump wave ($\lambda_p$) to achieve the enhanced SPDC process should be O-polarized. In addition, the generated signal and idler stream of photons are largely concentrated in the normal direction with respect to the metasurface. Indeed, Fig. 7 shows the photon-pair generation rate under different observation angles ($\theta_s$ and $\theta_i$) for the signal and idler waves, where $\theta_s$ and $\theta_i$ are defined with respect to the metasurface's normal direction. The used wavelengths are $\lambda_p = 679$ nm, $\lambda_s = 1257$ nm, and $\lambda_i = 1477$ nm. It is obvious that the radiated signal and idler photon power is restricted in a solid angle of few degrees around the normal direction. In addition, it is interesting that the direction with the maximum radiation power is not precisely normal to the sample surface. This can be explained again by studying the classical counterpart of the SFG process, where Figs. 7b and c show the maximum field enhancement in the LiNbO$_3$ layer as a function of the two pump wave incident angles. It can be seen that the peak field enhancement occurs when the input waves are slightly off the normal incidence. The incident angles ($\theta_1$ and $\theta_2$) of the peak field enhancement in the classical SFG process directly correspond to the observation angles ($\theta_s$ and $\theta_i$) of the photon-pair generation in the quantum SPDC process.

## 5. Conclusions

To conclude, a compact nonlinear plasmonic metasurface is proposed to efficiently generate entangled and correlated photon-pairs. By coinciding the metasurface fundamental and higher-order resonances with both the generated signal and idler frequencies, the electric field in the nonlinear media is significantly enhanced leading to a substantially boosted classical SFG nonlinear process that directly corresponds to an efficient quantum SPDC process. It is demonstrated that the metasurface design can effectively improve the efficiency of both SFG and SPDC processes due to strongly enhanced nonlinear light-matter interactions. The LiNbO$_3$ layer consists the dominant contribution in the second-order nonlinear process. The SPDC emitted streams of photons are highly directional and perpendicular to the metasurface plane. This out-of-plane directionality property overcomes the in-plane emission limitation of all-dielectric nanostructures, mainly due to the crystalline properties of III−V semiconductor materials, leading to easier detection of the generated entangled photon-pairs. The effect of polarization is also investigated because of the anisotropic nature of the presented second-order nonlinear response. The proposed free-standing SPDC source provides high efficiency, room temperature operation, out-of-plane directionality, and tunable response that can be further altered by changing the metasurface geometry or using different input and output wavelengths. Novel versatile quantum optical devices are envisioned based on the presented metasurface design that can be especially useful for free-space quantum plasmonic and other nanophotonic applications.

**Acknowledgements**
This work was partially supported by the NSF Nebraska Materials Research Science and Engineering Center (Grant No. DMR-1420645), Office of Naval Research Young Investigator Program (ONR-YIP) (Grant No. N00014-19-1-2384), National Science Foundation/EPSCoR RII Track-1: Emergent Quantum Materials and Technologies (EQUATE) under (Grant No. OIA-2044049), and the Jane Robertson Layman Fund from the University of Nebraska Foundation.

**Figures**

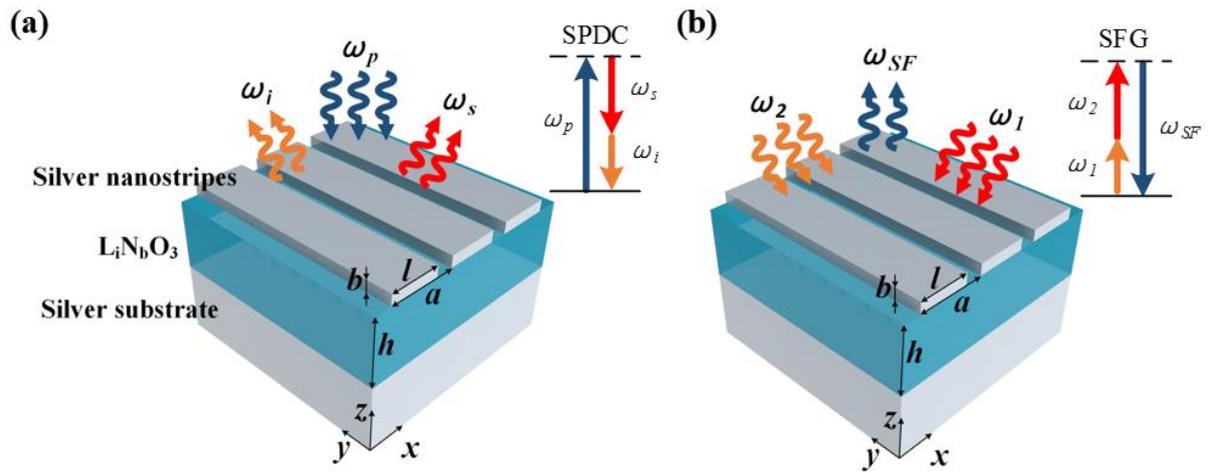

Fig. 1 Schematics of the proposed nonlinear plasmonic metasurface made of silver nanostripes, lithium niobate spacer layer, and silver substrate. The metasurface can boost both (a) quantum SPDC and (b) classical SFG processes, which are schematically shown in the insets of (a) and (b), respectively.

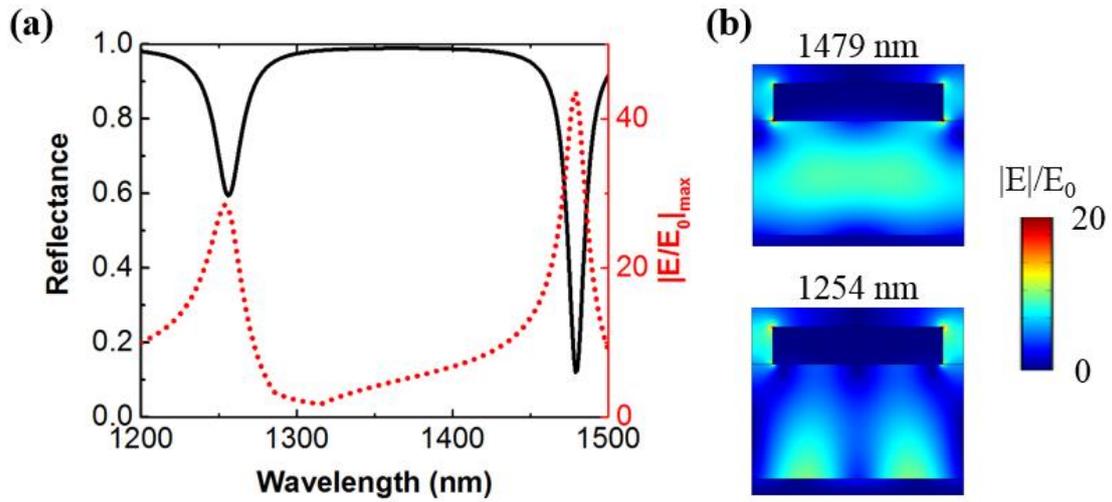

Fig. 2 (a) Linear reflectance and the maximum electric field enhancement in the LiNbO$_3$ layer as functions of the wavelength. (b) Electric field enhancement distribution at the fundamental (up) and higher-order (bottom) resonances.

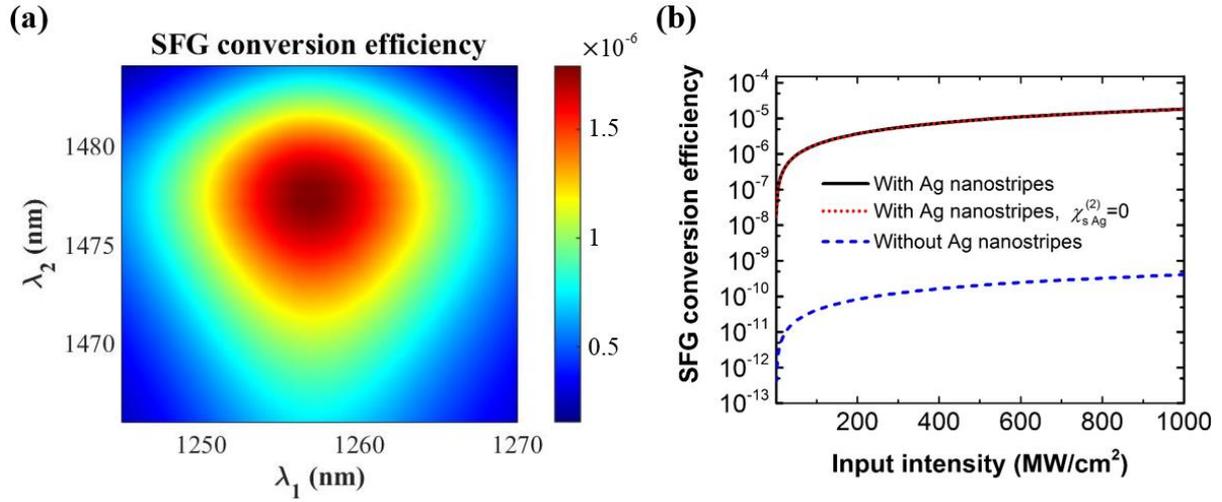

Fig. 3 (a) SFG conversion efficiency as a function of the two pump waves incident wavelengths. (b) SFG conversion efficiency as a function of the pump wave incident intensity for three different cases: i) all nonlinear materials included (black), ii) silver-dielectric interface nonlinearity removed (red), and iii) silver nanostripes removed from the metasurface design (blue).

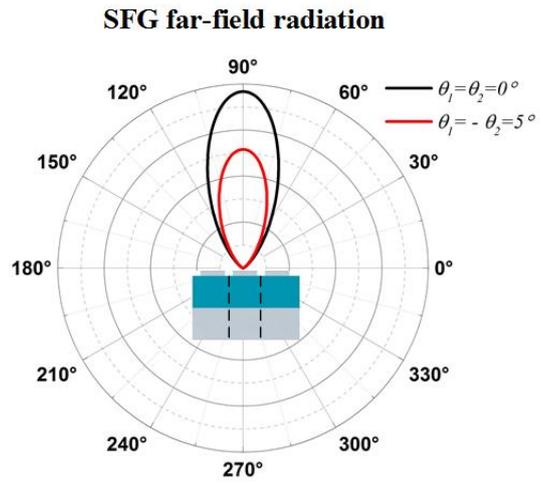

Fig. 4 SFG far-field radiation pattern from the metasurface under normal ($\theta_1=\theta_2=0°$) and oblique ($\theta_1= -\theta_2=5°$) incidence illumination.

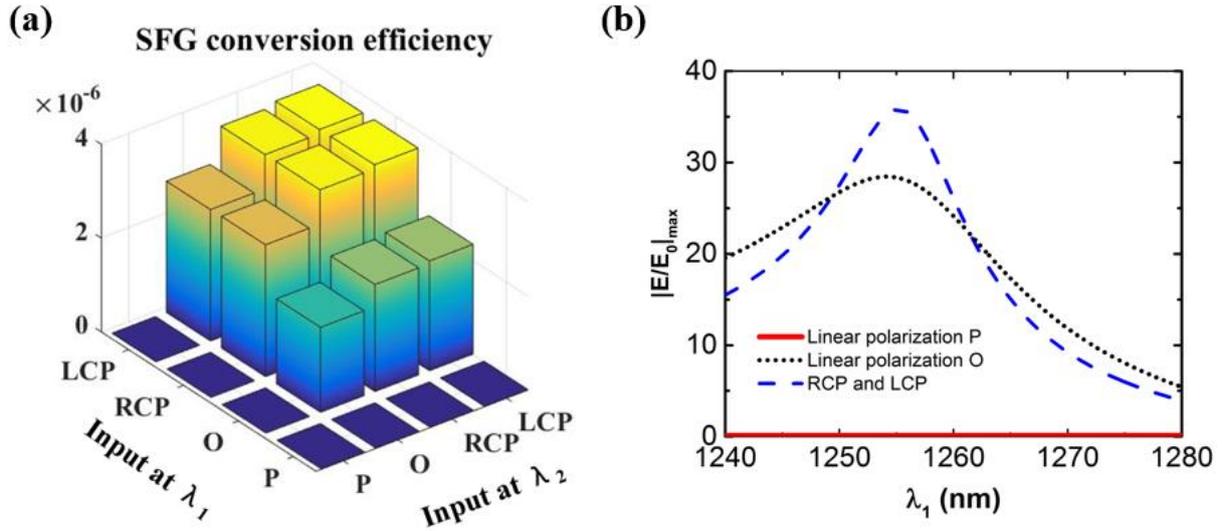

Fig. 5 (a) SFG conversion efficiency under different combinations of the input pump wave polarizations; O and P: linearly polarized with electric field $\mathbf{E}_0$ orthogonal and parallel to the silver nanostripes, respectively; LCP and RCP: left- and right-handed circularly polarized, respectively. (b) The maximum field enhancement in the LiNbO$_3$ layer as a function of the pump wavelength $\lambda_1$ for different polarizations.

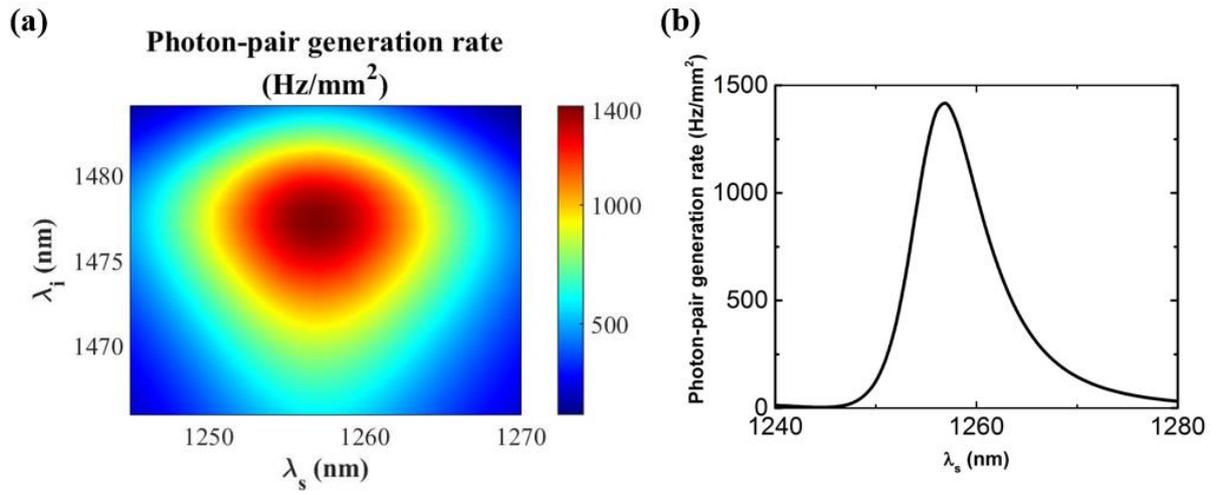

Fig. 6 (a) Photon-pair generation rate of the proposed nonlinear plasmonic metasurface as a function of the signal and idler wavelengths. (b) Photon-pair generation rate as a function of the signal wavelength when the pump wavelength is kept constant to $\lambda_p = 679$ nm.

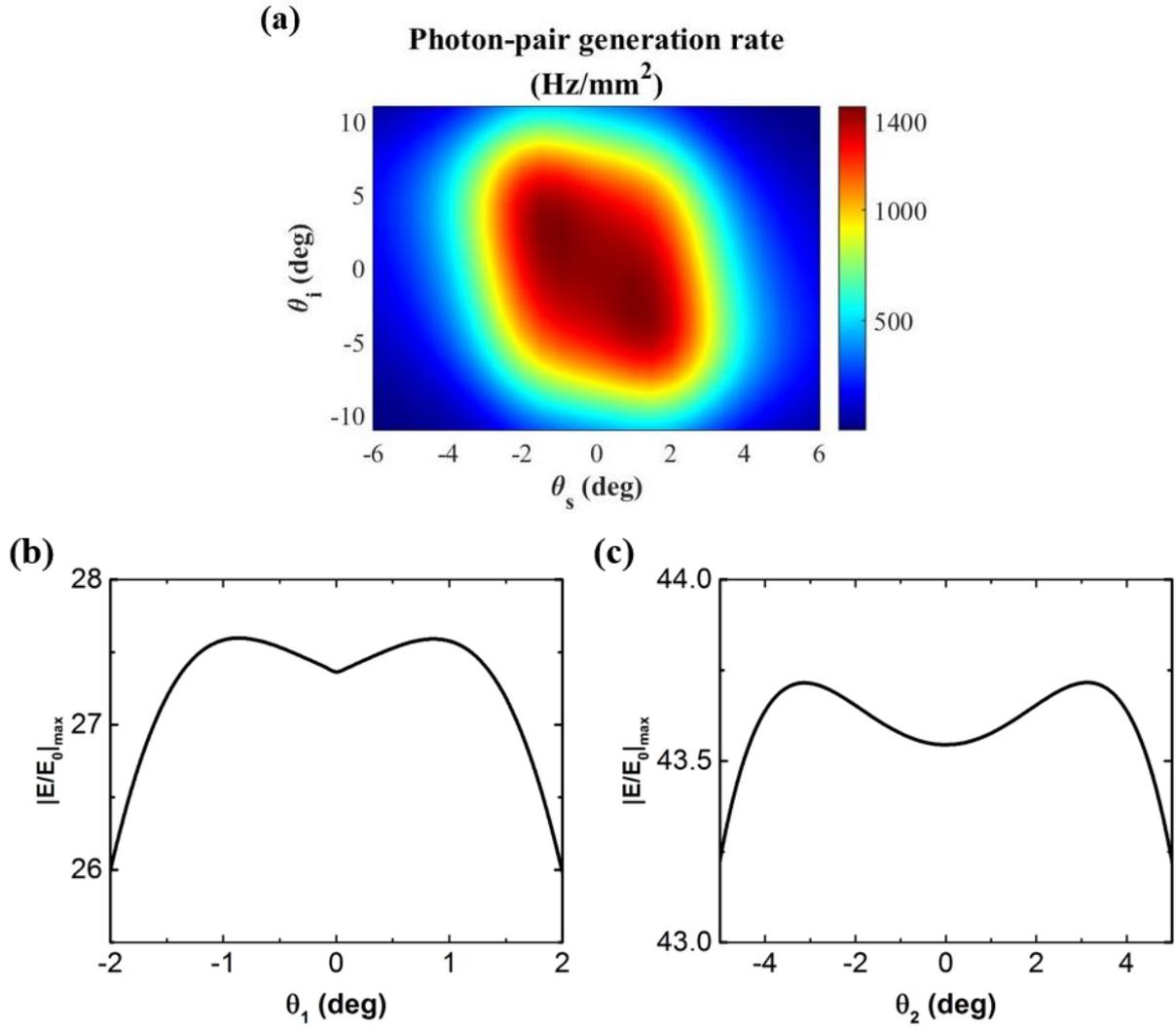

Fig. 7 (a) Photon-pair generation rate due to the boosted SPDC process as a function of the signal and idler wave observation angles ($\theta_s$ and $\theta_i$). (b-c) The maximum electric field enhancement in the LiNbO$_3$ layer as a function of (b) $\theta_1$ and (c) $\theta_2$ in the classical counterpart of the SPDC process, i.e., SFG.

# Supplementary Material

# Efficient single-photon pair generation by spontaneous parametric down-conversion in nonlinear plasmonic metasurfaces

Boyuan Jin, Dhananjay Mishra, and Christos Argyropoulos[*]

Department of Electrical and Computer Engineering, University of Nebraska-Lincoln, Lincoln, Nebraska 68588, USA

*christos.argyropoulos@unl.edu

1. **Numerical method to simulate the linear response and nonlinear SFG process**

In this work, the linear and nonlinear SFG processes are numerically modeled in the frequency domain by using COMSOL Multiphysics. COMSOL is a commercial simulation software based on the finite element method (FEM) and its RF module is used to carry out the presented full-wave electromagnetic simulations. These simulations are not trivial, since the COMSOL solver equations need to be substantially modified and customized to introduce the nonlinear response of the bulky material $LiNbO_3$ and metal-dielectric interface. More specifically, the second-order nonlinear response at the metal-dielectric interface is simplified and converted to a nonlinear surface current to reduce the computation load and accelerate the nonlinear simulations. More details about this part of the modeling will be provided later in this section. Similar simulation methods had been widely applied to numerically model various nonlinear optical effects, including SFG, based on plasmonic or dielectric structures and has been verified to accurately predict relevant various experimentally observed results.[1-5]

The proposed plasmonic metasurface (shown in Fig. 1 in the main paper) is composed of an array of silver nanostripes periodically placed on top of a dielectric spacer layer made of $LiNbO_3$. At the bottom, the substrate made of silver works as a perfect reflector and thus the plasmonic metasurface is working in the reflection mode. The silver nanostripes and substrate enhance the electric field distribution in the $LiNbO_3$ layer, as shown in Fig. 2b in the main paper. The proposed structures are uniform in the y-axis along the nanostripe direction, and therefore are simulated as two-dimensional (2D) systems. We also tried three-dimensional (3D) relevant simulations (not shown here), where it was verified that the results of the 2D model have great accuracy, but its calculations are less time consuming. Two ports are placed on the upper and bottom boundaries of the simulation domain operating at both incident wavelengths $\lambda_1$ and $\lambda_2$. Due to the periodicity of the structure, only one silver nanostripe needs to be included in the simulation domain and periodic boundary conditions are used in the left and right boundaries. The resulted input radiation is a typical plane wave source that propagates from the upper to the bottom port.

The linear permittivities of silver and LiNbO$_3$ are taken from experimental data.[6, 7]

During the SFG process in any nonlinear material, the incident waves operating at $\lambda_1$ and $\lambda_2$ can generate a new wave at the sum-frequency equal to $\lambda_{SF}$. The sum-frequency wave is generated by the nonlinear materials composing the relevant structure. Hence, the upper and bottom boundaries of the simulation domain are replaced by "passive" scattering boundaries to detect the generated wave at $\lambda_{SF}$. The nonlinear media in the plasmonic metasurface includes the LiNbO$_3$ dielectric layer and the silver-dielectric interfaces. The wave equation in frequency domain derived from Maxwell's equations is modified in the case of nonlinear simulations to:

$$\nabla \times (\mu_r^{-1} \nabla \times \mathbf{E}) - \varepsilon_r k_0^2 \mathbf{E} = \mu_0 \omega^2 \mathbf{P}^{NL}. \qquad (1)$$

The non-zero term $\mu_0 \omega^2 \mathbf{P}^{NL}$ on the right side of this equation is added into COMSOL by using a weak-form partial differential equation (PDE) module, similar to previous simulations of various nonlinear metamaterials.[1-4, 8]

At the silver-dielectric interface, the second-order nonlinearity arises from the surface nonlinear susceptibilities: $\chi^{(2)}_{s\perp,Ag}$ and $\chi^{(2)}_{s\|,Ag}$, where $\chi^{(2)}_{s\|,Ag}$ is very weak and is neglected.[9-11] Computing the polarizability $\mathbf{P}^{NL}_{s\perp} = 2\varepsilon_0 \chi^{(2)}_{s\perp,Ag} E_{1\perp,Ag} E_{2\perp,Ag} \hat{\mathbf{r}}_\perp$ is tricky because COMSOL cannot deal with a surface current in the normal to surface direction. Many numerical methods have been implemented to overcome this difficulty, such as using the nonlinear Mie-type solutions,[12] the weak form of the differential equations,[13] and the surface integral method.[14] In this work, $\mathbf{P}^{NL}_{s\perp}$ is assumed to be equivalent to a surface magnetic current density given by the formula: $\mathbf{J}^{NL}_{m,s} = \hat{\mathbf{r}}_\perp \times (\nabla_\| P^{NL}_{s\perp}) / \varepsilon'$, similar to previous published works.[15] The surface magnetic current can be directly calculated by COMSOL, since it only contains tangential components. The x and z components of $\hat{\mathbf{r}}_\perp$ are denoted by $n_x$ and $n_z$, respectively. The "down" and "up" functions are used to express the electric field in the silver surface, rather than in the adjacent dielectric media. Finally, the "dtang" function is used to obtain the gradient along the tangential surface.

As the SFG is a very weak process and the conversion efficiency $CE_{SFG}$ is usually less than few percent, the power transferred to the sum-frequency is much lower than the incident wave powers at $\lambda_1$ and $\lambda_2$. Therefore, the undepleted-pump approximation is adopted throughout all our SFG calculations. The transmittance and reflectance at the incident wavelengths $\lambda_1$ and $\lambda_2$ are accurately computed by linear simulations without the need of adding extra nonlinear terms. Note that the computational burden and simulation time consumption is substantially lower in the case of linear modeling.

To calculate the SFG efficiencies $CE_{SFG}$ and $\eta_{SFG}$, we need to measure $P_{SF}$, which is the reflected power at the sum-frequency. $P_{SF}$ is computed by setting boundary probes on all the outer boundaries of the simulation domain and then integrating the power density outflow at the sum-frequency. Normal incident waves are used in this work and the incident powers are given by: $P_1 = I_1 a$ and $P_2 = I_2 a$, where $a$ is the periodicity of the plasmonic metasurface. The Far-Field Domain module of COMSOL is used to calculate the near-to-far-field transformation of the computed at the near-field reflected power $P_{SF}$ at the sum-frequency point. This calculation happens for one unit cell because the presented metasurface is periodic. The same far-field calculations have been used before to compute the directivity of a different nonlinear process (four-wave-mixing) boosted by a similar metasurface[16] with much smaller gap size. The resulted SFG radiation pattern of the metasurface is shown in Fig. 4 in the main paper, where it is proven that the optimum SFG performance occurs only for normal incidence illuminations.

Finally, very fine mesh is used in the simulation domain, especially in the LiNbO$_3$ layer, and at the corners and edges of the silver nanostripe. The minimum mesh size is equal to 0.5 nm < $10^{-3} \lambda_{SF}$. This fine mesh guarantees the accuracy of the nonlinear simulation results and can accurately deal with potential instabilities caused by the enhanced electric field in the plasmonic metasurface.

## 2. Plasmonic metasurface excited by circular polarized incident waves

In Fig. 5b in the main text, we compare the field enhancement between the linear and circular polarizations. The incident intensity in Fig. 5b is chosen to be very low and the plasmonic metasurface operation is in the linear regime. Here, the reflectance spectra induced by circular polarized excitations are also computed and shown in Fig. S1. Note that the reflectance spectrum under the O-polarized excitation is shown in Fig. 2a in the main text. Both LCP and RCP incident waves can generate a deeper reflectance dip accompanied by a slightly narrower bandwidth. On the other hand, the resonant wavelengths are almost unchanged to those in the case of linear polarization, suggesting the proposed plasmonic metasurface will enhance the nonlinear process in a similar way for circular polarized inputs, as was shown in the main text. Finally, it is worth noting that the reflection response is the same for both LCP and RCP incident waves since the structure is not chiral.

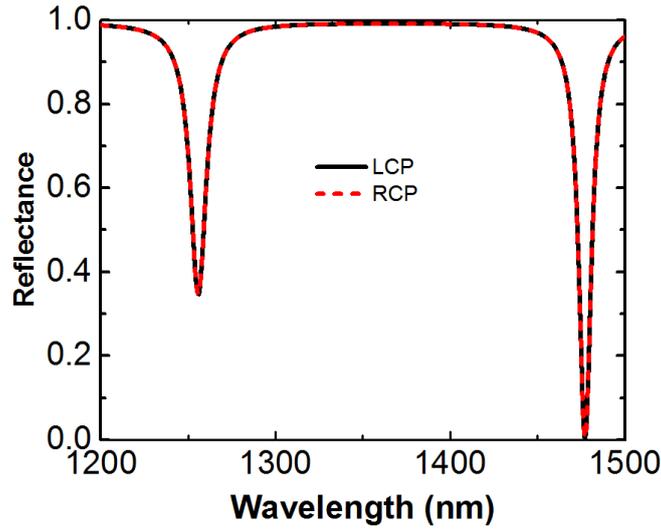

Fig. S1 Linear reflectance as a function of the wavelength when the incident light is circular polarized.

Interestingly, the reflectance is narrower and deeper in the case of circular polarization, as it is shown in Fig. S1. In addition, the field enhancement is higher in this case, as it is depicted in Fig. 5(b) in the main paper. This is the main cause of higher SFG conversion efficiency in the case of circular polarized incident waves with relevant results demonstrated in Fig. 5(a) in the main paper. Hence, it seems that stronger coupling exists between circular polarized excitation and the resonance modes of the presented plasmonic metasurface. In order to further understand this effect, the magnetic field distributions induced in the nanogap of the metasurface at the two reflectance resonances (1254 nm and 1479 nm) are plotted in Fig. S2, where it is demonstrated that both modes are of the same magnetic nature. Note that the electric field distributions of the same modes are shown in Fig. 2(b) in the main paper. It is interesting that the magnetic field is localized on the upper nanogap region in the 1479 nm resonance and on the lower nanogap region in the 1254 nm resonance. Magnetic modes are based on circulating electric fields and, as a result, can couple in a more efficient way to circular polarization, which is also the case in our design. Similar conclusions were derived in previous works that used the currently presented gap-plasmon metasurface mainly as polarizer.[17-20]

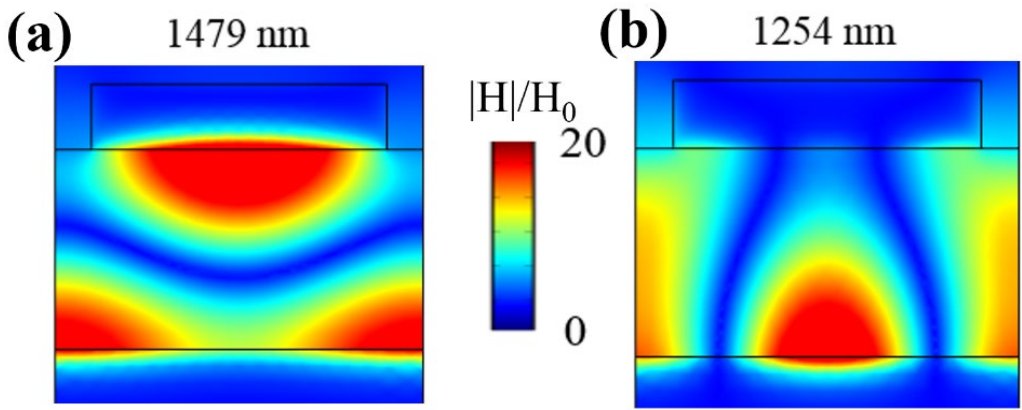

Fig. S2 Magnetic field enhancement distribution at the (a) fundamental and (b) higher-order resonances.